\documentclass[prb,twocolumn,showpacs,preprintnumbers,amsmath,amssymb,superscriptaddress]{revtex4}
\usepackage{graphicx}% Include figure files
\usepackage{dcolumn}% Align table columns on decimal point
\usepackage{bm}% bold math

\begin{document}

\preprint{APS/123-QED}

\title{Search for a spin--nematic phase in\\
the quasi-one-dimensional frustrated magnet LiCuVO$_4$}

\author{N. B\"{u}ttgen}
\email[]{e-mail: norbert.buettgen@physik.uni-augsburg.de}
%\thanks{}
\affiliation{Center for Electronic Correlations and Magnetism (EKM),
Experimentalphysik V, Universit\"{a}t Augsburg, D--86135 Augsburg, Germany}
\author{K. Nawa}
\affiliation{Department of Chemistry, Graduate School of Science, Kyoto University, Kyoto 606-8502, Japan}
\affiliation{Institute for Solid State Physics, The University of Tokyo, Kashiwanoha, Kashiwa, Chiba, 277-8581, Japan}
\author{T. Fujita}
\affiliation{Center for Advanced High Magnetic Field Science,
Graduate School of Science, Osaka University, Machikaneyama 1-1,
Toyonaka, Osaka 560-0043, Japan}
\author{M. Hagiwara}
\affiliation{Center for Advanced High Magnetic Field Science,
Graduate School of Science, Osaka University, Machikaneyama 1-1,
Toyonaka, Osaka 560-0043, Japan}
\author{P. Kuhns}
\affiliation{National High Magnetic Field Laboratory, Tallahassee, Florida 32310, USA}
\author{A. Prokofiev}
\affiliation{Institut f\"{u}r Festk\"{o}rperphysik Technische Universit\"{a}t Wien, A--1040 Wien, Austria}
\author{A.P. Reyes}
\affiliation{National High Magnetic Field Laboratory, Tallahassee, Florida 32310, USA}
\author{L.E. Svistov}
\email[]{svistov@kapitza.ras.ru} \affiliation{P.L. Kapitza Institute
for Physical Problems RAS, 119334 Moscow, Russia}
\author{K. Yoshimura}
\affiliation{Department of Chemistry, Graduate School of Science, Kyoto University, Kyoto 606-8502, Japan}
\author{M. Takigawa}
\email[]{masashi@issp.u-tokyo.ac.jp}
\affiliation{Institute for Solid State Physics, The University of Tokyo, Kashiwanoha, Kashiwa, Chiba, 277-8581, Japan}

\date{\today}% It is always \today, today,

\begin{abstract}
We have performed NMR experiments on the quasi one--dimensional frustrated spin-1/2
system LiCuVO$_4$ in magnetic fields $H$ applied along the $c$--axis up to
field values near the saturation field $H_{\rm sat}$. For the field range $H_{\rm c2}<H<H_{\rm c3}$
($\mu_0H_{\rm c2}\approx 7.5$~T and $\mu_0H_{\rm c3} = [40.5 \pm 0.2]$~T)
the $^{51}$V NMR spectra at $T$ = 380~mK exhibit a characteristic double--horn pattern,
as expected for a spin--modulated phase
in which the magnetic moments of Cu$^{2+}$ ions are aligned parallel
to the applied field $H$ and their magnitudes change sinusoidally along the magnetic chains.
For higher fields, the $^{51}$V NMR spectral shape changes from the double--horn pattern
into a single Lorentzian line. For this Lorentzian line, the internal field at the $^{51}$V nuclei stays constant for
$\mu_0 H > 41.4$~T, indicating that the majority of magnetic moments in LiCuVO$_4$
are already saturated in this field range. This result is inconsistent with the previously
observed linear field dependence of the magnetization $M(H)$ for $H_{\rm c3}<H<H_{\rm sat}$ with $\mu_0H_{\rm sat}=45$~T
[L. E. Svistov {\it et al}., JETP Letters {\bf 93}, 21 (2011)]. We argue that the discrepancy is due
to non-magnetic defects in the samples.  The results of the spin--lattice relaxation rate of $^7$Li
nuclei indicate an energy gap which grows with field twice as fast as the Zeeman
energy of a single spin, therefore, suggesting that the two--magnon bound state is the lowest energy excitation.
The energy gap tends to close at $\mu_0H \approx 41$~T. Our results suggest that the
theoretically predicted spin--nematic phase, if it exists in LiCuVO$_4$,
can be established only within the narrow field range $40.5 < \mu_0 H < 41.4$~T .
\end{abstract}

\maketitle

\section{Introduction}
The search for novel quantum states in strongly correlated electronic
systems exhibiting an exotic order is at the core of modern
condensed matter physics. An example that attracts strong recent
interest is the \textit{nematic phase}. Analogous to the nematic
state observed in liquid crystals, electronic correlation in a
nematic phase develops a \textit{preferred orientation}, which
breaks rotational symmetry. However, time reversal symmetry is
preserved unlike in a conventional magnetic order. The possibility of
such nematic order has been discussed in several different systems
such as frustrated magnets,\cite{Andreev_84,Penc_11} iron--pnictide
superconductors,\cite{Kasahara_12} and heavy--fermion
materials.\cite{Okazaki_11}

Among them, one--dimensional (1D) frustrated quantum spin systems ($S=1/2$) having a
ferromagnetic nearest neighbor (NN) exchange interaction $J_1$ and
an antiferromagnetic next--nearest neighbor (NNN) interaction $J_2$,
\begin{equation}
\label{J1J2}
\mathcal{H} = J_1 \sum _{n} \mathbf{S}_n \cdot \mathbf{S}_{n+1} +
J_2 \sum _{n} \mathbf{S}_n \cdot \mathbf{S}_{n+2}
\end{equation}
in magnetic fields deserve particular attention since extensive theoretical efforts
have been devoted to this
model.\cite{Kecke_07,Vekua_07,Hikihara_08,Sudan_09,Heidrich_09,Sato_09,Sato_11,Zhitomirsky_10,Nishimoto_12,
Ueda_09,Sato_13,Starykh_14,Ueda_14} Such intrachain interactions together with ferromagnetic interchain coupling
in the case of large spin values yield a rich magnetic phase diagram.\cite{Nagamiya_62} In particulary, the case $| J_1/(4J_2) | < 1 $ leads
to a helical magnetically ordered structure. In principle, a true long--range magnetically ordered structure
of a 1D chain of magnetic moments is suppressed by spin fluctuations, and a magnetic
state of such a system is characterized by short--range correlations between the magnetic moments.
But according to theory, a long--range magnetic order can be stabilized due to interactions beyond Eq.~(\ref{J1J2}), i.e.,
interactions with an applied magnetic field, interactions with the crystalline environment, or interactions between magnetic chains.
For instance a moderate static magnetic field $H$ stabilizes helical correlations in favour of a long--range ordered state even in one dimension\cite{Hikihara_08,Sudan_09}
with the order parameter $<\mathbf{S}_n \times
\mathbf{S}_{n+1}>_z = <S^x_{n}S^y_{n+1}-S^y_{n}S^x_{n+1}>$ for $S=1/2$ and $H \parallel z$. In contrast to the helical magnetic structure
the expectation values of the transverse spin components $<S^x_{n}>$ and $<S^y_{n}>$ are equal to zero. Such magnetic state with this pseudovector
order parameter was labeled the chiral phase in the literature. To our knowledge, this chiral phase has not been observed experimentally.

One--dimensional frustrated quantum spin systems with competing exchange interactions given by Eq.~(\ref{J1J2}) are
intriguing also in high magnetic fields, near the saturation field $H_{\rm sat}$. Let us first consider the fully
polarized ground state above the saturation field $H > H_{\rm sat}$.  In ordinary unfrustrated antiferromagnets, the lowest energy excitation
is described by a single spin--flip, or a magnon, which acquires kinetic energy
due to exchange interactions, forming a dispersive magnon band. As the field is reduced,
the excitation gap determined by the minimum energy of the magnon band vanishes at $H_{\rm sat}$.
Below $H_{\rm sat}$, the Bose-Einstein condensation (BEC) of magnons
occurs in the presence of finite three--dimensional interchain coupling, resulting in an antiferromagnetic
order in the plane perpendicular to the field with the order parameter $(-1)^n <S^-_n  + H.c.>$
(Ref. \onlinecite{Matsubara_56,Batyev_84}).

In contrast, the lowest energy excitation in the frustrated chain of
Eq.~(\ref{J1J2}) immediately above $H_{\rm sat}$ is a bound magnon
pair, not a single spin-flip, stabilized by the ferromagnetic NN
interaction for a wide range of the ratio $J_1/J_2$ (Refs.~\onlinecite{Chubukov_91,Kuzian_07,Kecke_07,Vekua_07,Hikihara_08,Sudan_09,Heidrich_09,Shannon_13}).
Below $H_{\rm sat}$, these bound magnon pairs exhibit BEC in the presence of special--type interchain coupling and establish a
long--range ordered magnetic structure with the order parameter given by
$(-1)^n <S^-_nS^-_{n+1} + H.c.> =
(-1)^n<(S^x_{n}S^x_{n+1}-S^y_{n}S^y_{n+1})>$ (Refs.~\onlinecite{Zhitomirsky_10,Nishimoto_12,
Ueda_09,Sato_13,Starykh_14,Ueda_14}). This type of magnetic order in the recent literature is labeled as {\em nematic}.
Again, the nematic order breaks the spin rotational symmetry. More precisely, the preferred orientation of the nearest neighbor spin correlation (director) rotates $90^\circ$ from one bond to the next. However, the transverse components $<S^x_{n}>$ and $<S^y_{n}>$ are zero and
the longitudinal magnetization remains uniform, $<S^z_{n}>=M/N$, where $M$ is the bulk magnetization and $N$ is the
number of magnetic ions of the sample.

As the field is further decreased, theories predict that the nematic
order is replaced by a spin--density--wave (SDW) state, where the
moments are collinear with the external field and their magnitudes
are modulated with a generally incommensurate
periodicity.\cite{Hikihara_08,Sudan_09,Heidrich_09,Ueda_09,Sato_13,Starykh_14}
This SDW state can be regarded as a spatial order of the bound
magnon pairs stabilized by their mutual repulsive interaction, which
becomes dominant over the gain in the kinetic energy of the BEC
phase as the density of bound magnon pairs increases.

\begin{figure}
\includegraphics[width=85 mm,angle=0,clip]{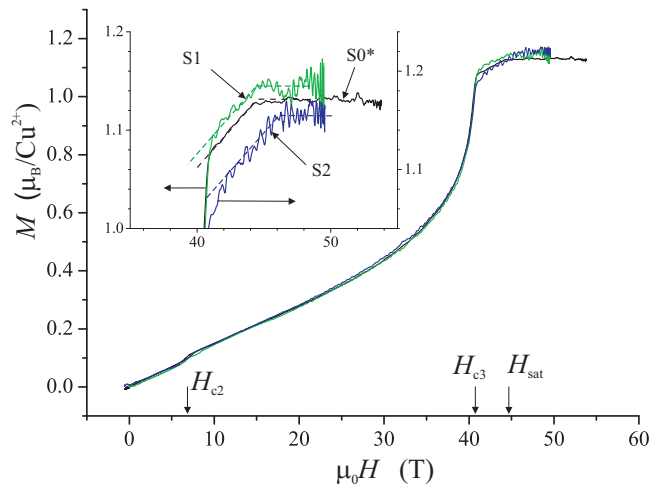}
\caption{Magnetization $M(H)$ measured in pulsed magnetic fields $H$
oriented $\mathbf{H}\parallel \mathbf{c}$ at $T = 1.3$~K in
LiCuVO$_4$. Inset: Magnification of $M(H)$ for fields around the
saturation field $H_{\rm sat}$. Data of three different single
crystals S1, S2 and S0$^*$ are shown, where the data of S0$^*$ are
taken from Ref. \onlinecite{Svistov_11}. The dashed lines are
parallel to each other and guide the eye. The $M(H)$ data for the
samples S1 and S2 are more noisy than for S0$^*$, because these
samples are significantly smaller.} \label{Figure-1}
\end{figure}

A promising candidate compound where these theoretical
predictions are expected to be tested experimentally is the antiferromagnet
LiCuVO$_4$, which contains frustrated chains of Cu$^{2+}$ ($S=1/2$)
along the crystallographic $b$--direction.\cite{Prokofiev_04}
Let us first summarize the results of the previous studies on the
magnetic phases of LiCuVO$_4$ for fields lower than
$\mu_0 H = 30$~T (Refs. \onlinecite{Gibson_04,Enderle_05,Kegler_06,Buettgen_07,Buettgen_10,Buettgen_12,Nawa_13,Enderle_12,Masuda_11}).
At zero field, an incommensurate planar spiral structure is realized
below $T_{\rm N}\approx 2.3 $~K with the moments lying in the
$ab$--plane.\cite{Gibson_04,Enderle_05} A spin--flip transition occurs at $\mu_0 H_{\rm c1} = 2.5$~T,
flipping the moments into the plane perpendicular to the field.\cite{Buettgen_07}
In higher fields $H > H_{\rm c2}$ ($\mu_0H_{\rm c2} \approx 7.5$~T) a
collinear spin--modulated structure is
realized,\cite{Buettgen_07,Buettgen_10,Buettgen_12,Nawa_13,Enderle_12}
consistent with the theoretical prediction.\cite{Hikihara_08,Sudan_09,Heidrich_09,Ueda_09,Sato_13,Starykh_14}
Remarkably, the relation between the magnetization and the wave vector of the spin--modulated
structure measured by neutron diffraction experiments in fields up
to 15~T demonstrates that it is the bound magnon pairs with $S_z =
2$ that form a periodic structure.\cite{Masuda_11,Enderle_12}  Also
the temperature dependence of the nuclear spin--lattice relaxation
rate $1/T_1$ at the V sites indicates the development of an energy gap in
the transverse spin excitation spectrum above $T_{\rm
N}$,\cite{Nawa_13} which corresponds to the binding energy of magnon
pairs, as theoretically predicted.\cite{Sato_09,Sato_11} The
observation of the spin--modulated phase and the experimental proof
for the bound magnon pairs provide strong support that LiCuVO$_4$ is
indeed well described by the model of Eq.~(\ref{J1J2}).

By further increase of the applied magnetic field $H$, it is expected
that the spin--nematic phase develops before the magnetization
saturates at $H_{\rm sat}$. In recent experiments the field
dependent magnetization curve $M(H)$ of LiCuVO$_4$ exhibited
anomalies slightly below $\mu_0H_{\rm sat} \approx 45$~T, indicating a new
phase.\cite{Svistov_11} Figure 1 shows $M(H)$ for three different samples
measured in pulsed magnetic fields $H$ applied along the $c$--direction at $T = 1.3$~K.
In addition to the anomalies at $H_{\rm c2}$ and $H_{\rm sat}$ already mentioned,
all samples show another anomaly at $\mu_0H_{\rm c3} \approx 40.5$~T,
about 5~T below the saturation.  The sharp increase of the magnetization
towards higher field stops at $\mu_0H_{\rm c3}$. The magnetization then
increases linearly in the field range $H_{\rm c3} \lesssim H < H_{\rm sat}$
with the slope $\approx 1/2 \cdot M_{\rm sat}/H_{\rm sat}$.
This magnetization behavior near saturation including the slope of $M(H)$
for $H_{\rm c3} \lesssim H < H_{\rm sat}$ is very similar for all sample.

In this paper we discuss the NMR spectra of $^{51}$V nuclei and the
nuclear spin--lattice relaxation rate $1/T_1$ of $^{7}$Li nuclei for
applied magnetic fields $H$ along the $c$--direction
($\mathbf{H}\parallel \mathbf{c}$) near the saturation field $\mu_0
H_{\rm sat} \approx 45$~T. The NMR spectra represent distributions
of the internal magnetic fields which are produced by the
surrounding magnetic Cu$^{2+}$ moments. The internal field has a
large distribution in the spin--modulated phase leading to broad NMR
spectra. In the nematic phase, in contrast, moments are uniform and
we expect a sharp single, solitary NMR line similar to the saturated
phase. However, unlike in the saturated phase, the internal field
determined from the peak position of the spectrum should change with
the magnetic field $H$ as it is the case for the bulk magnetization $M(H)$ in the
nematic phase. Such expectation may not be realized if the sample contains defects,
which generate inhomogeneous distribution of the magnetic moments. Since
the NMR spectra reflect the histogram of magnetic moments while
$M(H)$ gives average magnetization, they may behave differently as we discuss below.

\section{Experimental Results}
The observation of the phase transition from the spin-modulated
phase to the magnetically saturated phase from a local point of view
is the major subject of the present work. In addition to $M(H)$
measurements, which give bulk properties, we employed NMR of
$^{51}$V ($I=7/2$, $\gamma = 2\pi \times 11.1988$~MHz/T) nuclei
probing the local magnetic properties of the Cu$^{2+}$ moments in
LiCuVO$_4$. Three  single--crystalline samples were used in the NMR
measurements: sample S0, S1, and S2. Sample S0 was used in the previous NMR experiments
in Refs.~\onlinecite{Buettgen_07,Buettgen_10,Buettgen_12} and obtained from the same batch
as S0$^*$ whose $M(H)$ curve is shown in Fig. \ref{Figure-1}. Samples S1 and S2 are from a new batch
whose $M(H)$ curves are also shown in Fig. \ref{Figure-1}. Spin--echo techniques were utilized in the hybrid 45~T
magnet of the DC field facility at the National High Magnetic Field Laboratory (NHMFL), Tallahassee, Florida. The absolute
values of the applied magnetic fields $H$ were calibrated by NMR using
aluminum powder. All spectra of sample S0 were collected with the
pulse sequence $3\mu s -\tau -3\mu s$ ($\tau =15\mu s$) by sweeping
the applied magnetic field $H$ at constant frequencies. Spectra of
the single--crystalline samples S1 and S2 were recorded by summing
the Fourier transform of the spin-echo signal with the pulse
sequence $2.5\mu s -\tau -2\mu s$ ($\tau = 17\mu s$) obtained for
equally spaced excitation frequencies at a fixed magnetic field.
Temperatures down to $T = 380$~mK were achieved within a $^3$He
cryostat.

\begin{figure}
\includegraphics[width=85 mm]{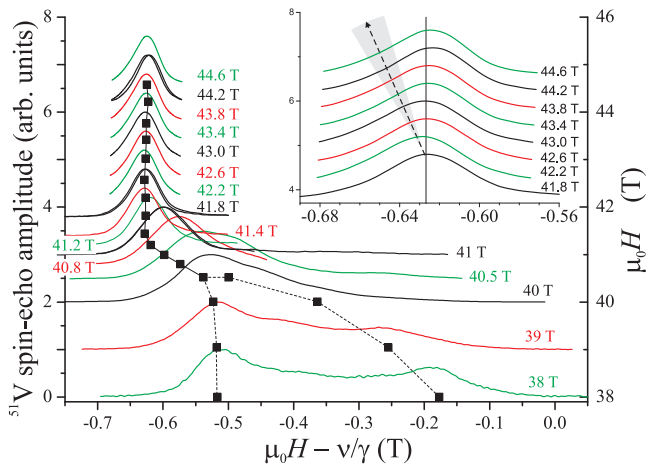}
\caption{$^{51}$V NMR spectra obtained from crystal S2 at
$T = 380$~mK within the field range $38 \le \mu_0 H \le 45$~T
for $\mathbf{H}\parallel \mathbf{c}$. The spin-echo amplitudes are normalized by the
peak intensity. The spectral shape shows a crossover at $\mu_0H_{\rm c3} \approx 40.5$~T
from the double-horn pattern at lower fields to the single-peak pattern
at higher fields. The dashed arrow in the inset denotes
the expected line shift which is estimated from the bulk
magnetization $M(H)$.}
 \label{Figure-2}
\end{figure}

\begin{figure}
\includegraphics[width=85 mm]{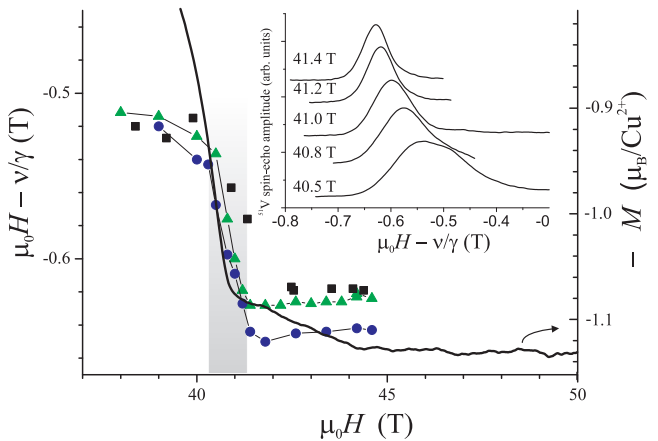}
\caption{Peak positions of the $^{51}$V NMR spectra in LiCuVO$_4$
are plotted against applied field along the $c$--direction obtained
from samples S0 (squares), S1 (triangles), and S2 (bullets),
respectively (left axis). The solid line shows the bulk
magnetization $M(H)$ vs. the applied magnetic field $H$ obtained
from sample S0$^*$ (right axis). Inset: $^{51}$V NMR spectra of
sample S2.} \label{Figure-3}
\end{figure}

Figure \ref{Figure-2} shows the NMR spectra of sample S2 obtained at
$T = 380$~mK within the range $38 \le \mu_0 H \le 45$~T of the
applied magnetic field $H$ for $\mathbf{H}\parallel \mathbf{c}$.
Note that $\nu/\gamma \equiv \mu_0H_{\rm eff}$ represents the
effective magnetic field sensed by nuclei being observed at the
resonance frequency $\nu$ and is the sum of the applied field plus
the internal local field due to Cu$^{2+}$ moments, i.e., $H_{\rm eff} = H
+ H_{\rm int}$. Therefore, the abscissa axis of the spectra in
Fig.~\ref{Figure-2} $\mu_0H - \nu/\gamma$ is equal to $-\mu_0H_{\rm
int}$ at the probing $^{51}$V nuclei. The double-horn shaped pattern
for lower fields $H$ at the bottom of Fig.~\ref{Figure-2} resemble
our previous results,\cite{Buettgen_07,Buettgen_10,Buettgen_12} where we
established the spin-modulated magnetic structure in LiCuVO$_4$. The
asymmetry of the double--horn shaped pattern for $\mu_0 H < 30$ T
turned out to be due to different spin--spin relaxation times $T_2$
at different spectral positions.\cite{Buettgen_12} Around the value
of the applied magnetic field $\mu_0 H_{\rm c3} = 40.5$~T the
double--horn shape starts to disappear in favor of a single--peak
spectrum evolving a more symmetric line shape towards higher magnetic
fields $H$.

For applied magnetic fields $H$ higher than $\mu_0 H > 41.4$~T, the
internal field occurs to stay almost constant without any shift in
the peak position of the spectral line. This observation is in
contrast to the $M(H)$ data which further increase linearly for the
same field values up to the saturation field $H_{\rm sat}$ (cf. the
inset in Fig. \ref{Figure-1}). This conflict is illustrated in the
inset of Fig. \ref{Figure-2}, where the almost unshifted NMR
spectral lines are plotted together with the internal field fields
calculated from the $M(H)$ data (dashed arrow). The shaded sector in
the inset of Fig. \ref{Figure-2} shows the estimated error made for
the slope of the $M(H)$ dependence obtained from pulse field
experiments on $M(H)$ (cf. Fig. \ref{Figure-1}). Such unexpected
behavior was observed for all three studied samples. The field
dependences of the peak positions of the $^{51}$V NMR spectra are
shown in Fig. \ref{Figure-3} together with the $-M(H)$ data. The
internal field at the vanadium nuclei for all three samples shows
almost no change with field for $\mu_0 H> 41.4$ T in contrast to the
full magnetic moment of the sample $M(H)$ measured in the same field
range. This discrepancy between NMR line shift and bulk
magnetization $M(H)$ demonstrates that the magnetization is not
uniform, probably due to the presence of defects. The constant internal field at
the peak of the intense NMR line indicates that the majority part of the nuclei
far from defects is embedded in an electronic surrounding with saturated
magnetic moments, while the bulk magnetization is affected by defects. This
point is discussed in more detail below. Therefore,the NMR result is seemingly
ineffective in detecting the existence of a spin nematic phase expected within
the field range $H_{\rm c3} < H < H_{\rm sat}$ previously.

However, close inspection of the data in Fig.~\ref{Figure-3} reveals that within
a limited narrow field range $40.5 \le \mu_0 H \le 41.4$~T (indicated by the
shaded area in Fig.~\ref{Figure-3}) the NMR spectral pattern is
characterized by a nearly symmetric, solitary, single peak whose resonance position
shifts strongly depending on the applied magnetic field $H$ (cf. the inset of Fig.~\ref{Figure-3}).
This observation meets the theoretically predicted behavior of the spin--nematic phase.
Indeed, the behavior of the NMR internal field within the restricted field range mentioned above is
consistent with the theoretical prediction of a very steep slope of the magnetization curve in
the nematic phase immediately below saturation.\cite{Ueda_14}

\begin{figure}
\includegraphics[width=85 mm]{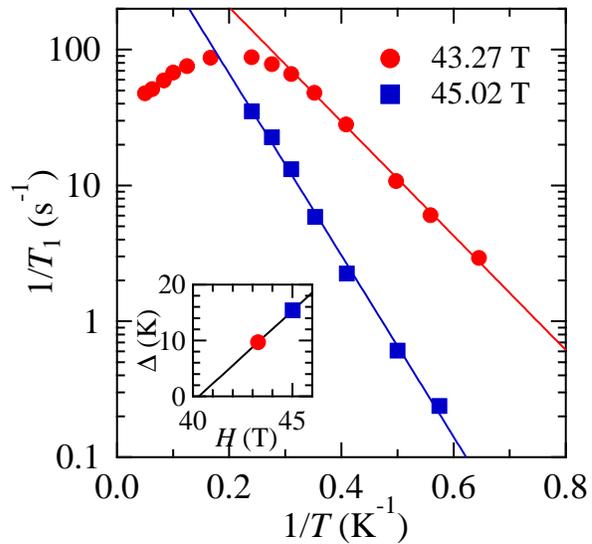}
\caption{Temperature dependence of the spin--lattice relaxation rate
$1/T_1$ at $^7$Li nuclei in Arrhenius representation measured for
different applied magnetic fields $\mu_0 H = 43.27$~T and 45.02~T. Inset:
The slopes of the Arrhenius representations, i.e. activation energies $\Delta$, plotted vs. the applied magnetic field $H$. The solid line indicates a linear fit (see text).
}
 \label{Figure-4}
\end{figure}

\begin{figure}
\includegraphics[width=85 mm]{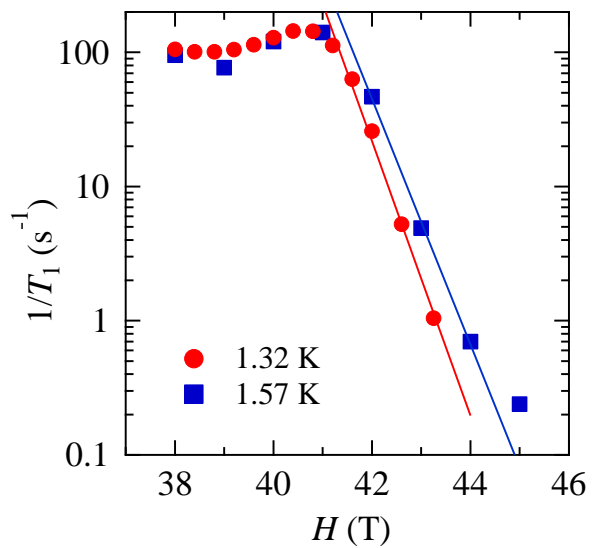}
\caption{Field dependence of the spin--lattice relaxation rate
$1/T_1$  at $^7$Li nuclei for different temperatures.}
 \label{Figure-5}
\end{figure}

After the description of static magnetic properties probed by $^{51}$V NMR mentioned above,
we now turn to dynamic properties and the results of the nuclear
spin--lattice relaxation rate $1/T_1$ of $^7$Li nuclei ($I$=3/2, $\gamma = 2\pi\times
16.5466$~MHz/T). The sample used for the $1/T_1$ measurements is
none of the three samples used for the $^{51}$V NMR, but is from the same
batch as the sample used in one of the previous NMR experiments
described in Ref.~\onlinecite{Nawa_13}. Figure \ref{Figure-4} shows
the temperature dependence of $1/T_1$  at two different applied
magnetic fields above $\mu_0 H = 41.4$~T. The Arrhenius
representation of the data exhibits a thermally activated behavior
at low temperatures
\begin{equation}
\label{activation}
\frac{1}{T_1} \propto \exp \left( - \frac{\Delta}{T} \right).
\end{equation}
The activation energies are obtained as $\Delta =15.4$~K at 45.02~T
and $\Delta = 9.7$~K at 43.27~T (see the inset of Fig.~\ref{Figure-4}).
Assuming that the activation gap depends linearly on the field $H$,
\begin{equation}
\label{gap}
\Delta=D(H-H_0),
\end{equation}
we obtain $D = 3.3$~K/T. In ordinary antiferromagnets, where the
lowest energy excitations in the saturated state are single
spin--flip processes involving the change of magnetization by
$\Delta S_z = 1$, the excitation gap should be given by $g_c\mu_{\rm
B} (H - H_{\rm sat})$ with $g_c\mu_{\rm B} = 1.55$~K/T ($g_c$ =
2.31, Ref.~\onlinecite{Nidda_02}). Remarkably, the value of $D$ is
much larger than $g_c\mu_{\rm B}$ but close to $2g_c\mu_{\rm B} =
3.1$~K/T. This indicates that the lowest--energy excitations are
bound magnon pairs with $\Delta S_z = 2$.

The assumption of the linear field dependence of the energy gap
is indeed justified by the exponential field dependence of $1/T_1$
above 41~T
\begin{equation}
\label{exp_field}
\frac{1}{T_1} \propto \exp \left( - A H \right)
\end{equation}
as shown in figure~\ref{Figure-5}. From the plots for two different
temperature values we obtain $A=2.34$~T$^{-1}$ for $T=1.32$~K and
$A=2.10$~T$^{-1}$ for $T=1.57$~K. By relating Eq.~(\ref{exp_field})
to Eqs.~(\ref{activation}) and (\ref{gap}), we obtain $D = 3.1$~K/T
(3.2~K/T) for $T=1.32$~K ($T=1.57$~K). These values are again close
to $2g_c\mu_{\rm B}$, consistent with the picture that the lowest
excitation is a two--magnon bound state.

From the plot in the inset of Fig.~\ref{Figure-4}, we expect the
excitation gap to become zero at applied magnetic fields $\mu_0 H\approx 41$~T.
This value is different from $\mu_0 H_{\rm sat} = 45$~T determined
from the magnetization data but agrees with the field value above
which the $^{51}$V NMR spectral peak stays constant. This fact
provides further support for our previous conclusion that a majority
of the magnetic copper ions in LiCuVO$_4$ are already saturated {\em below}
$\mu_0 H_{\rm sat} = 45$~T.

\section{Discussion}
\begin{figure}
\includegraphics[width=85 mm]{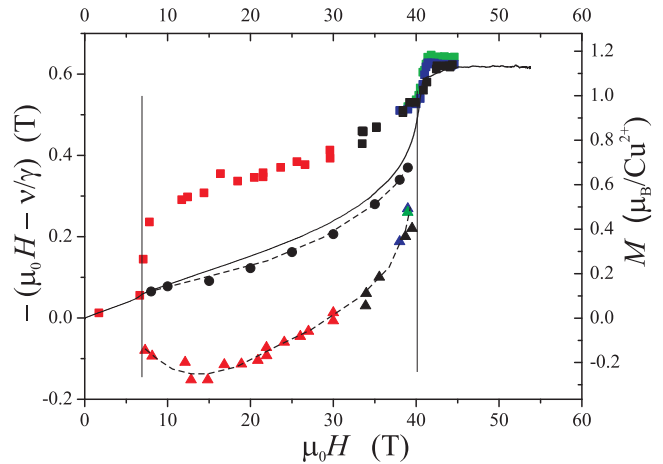}
\caption{Left axis: Field dependence of the $^{51}$V spin-echo
amplitude maxima of the double-horn pattern shown in Fig.
\ref{Figure-2} (red symbols are data taken from Ref.
\onlinecite{Buettgen_12}). The other symbols denote the data of
samples S0 (black), S1 (blue), and S2 (green), respectively. The
bullets denote the arithmetic average of high-field (triangles) and
low-field (squares) maxima. The dashed lines guide the eye. Right
axis: The black solid line is the magnetization measurement $M(H)$
of sample S0$^*$ taken from Ref. \onlinecite{Svistov_11}.}
 \label{Figure-6}
\end{figure}

We first discuss the $^{51}$V NMR spectra in the field range $H_{\rm
c2} < H < H_{\rm c3}$, where the magnetically ordered structure is
recognized as the spin--modulated structure. The characteristic
double--horn shape (cf. Fig.~\ref{Figure-2}) is an unambiguous
fingerprint of an incommensurate magnetic structure only observed
within this field range. For lower applied magnetic fields $H <
H_{\rm c2}$ or higher fields $H > H_{\rm c3}$ the $^{51}$V NMR
spectra consist of a single peak. The field dependence of the
internal fields at the spectral maxima of the low--field and
high--field horn in the case of the double-horn pattern, as well as the
internal field of the singly peaked spectra are plotted in Fig.~\ref{Figure-6}
(left axis). The arithmetically averaged field values
$H_{\rm max}$ (solid bullets) of the high--field (triangles) and
low--field (squares) maxima, together with the single--peak maxima
tightly follow the field dependence of the magnetization
data $M(H)$ (right axis in Fig. \ref{Figure-6}).

\begin{figure}
\includegraphics[width=80 mm]{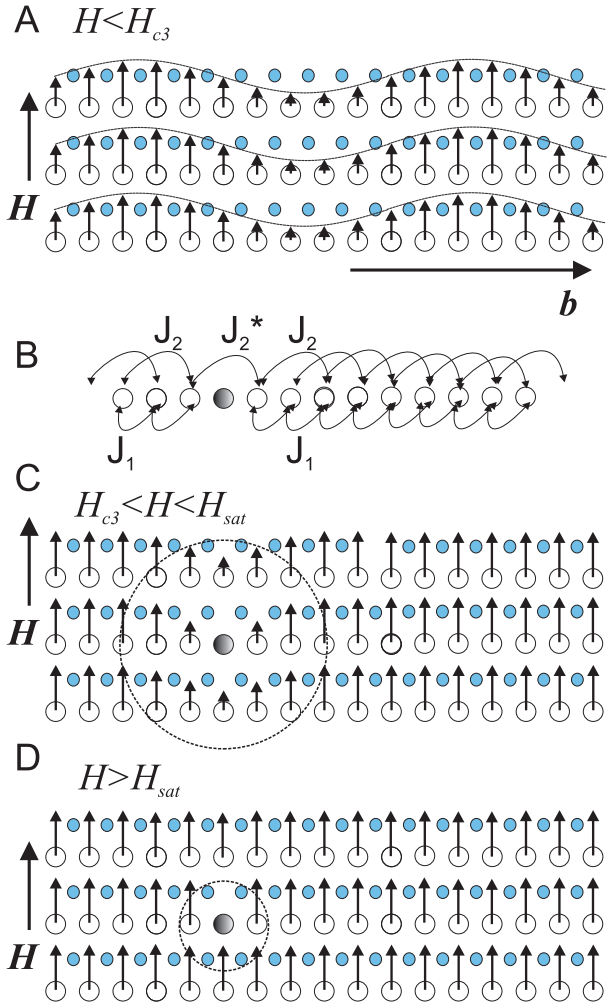}
\caption{Models of the magnetic structure projected along the
$a$-axis for different field ranges, illustrating Cu$^{2+}$ ions
(large open circles), V$^{5+}$ ions (small blue circles), and
magnetic moments of Cu$^{2+}$ (arrows). Panel A: the sinusoidal
modulation below $H_{\rm c3}$ reconstructed from the values of
$\mu_m$ and $\mu_1$ obtained by fitting the $^{51}$V NMR spectrum at
$\mu_0 H = 38$~T, just below $H_{\rm c3}$. Panel B: illustration of
intrachain exchange bonds with a defect replacing a magnetic ion by
a nonmagnetic one (gray bullet). Panels C and D: models of the
magnetic structure with a defect for fields $H_{\rm c3}<H<H_{\rm
sat}$ and $H>H_{\rm sat}$, respectively. The dashed circles around
the non--magnetic defect mark the region where the internal field
of $^{51}$V nuclei is smaller than the majority because the
neighboring Cu moments are unsaturated.} \label{Figure-7}
\end{figure}

\begin{figure}
\includegraphics[width=80 mm]{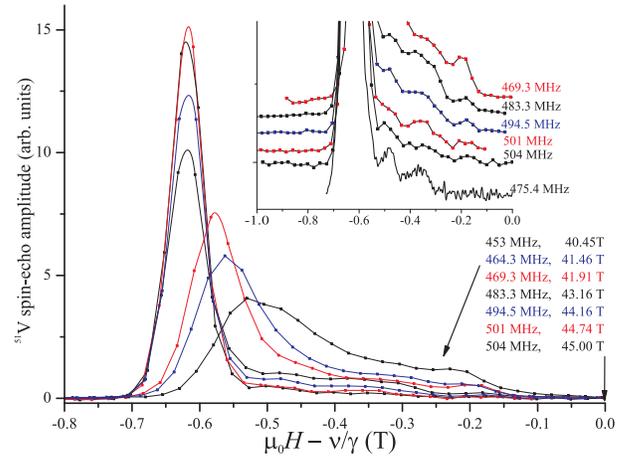}
\caption{$^{51}$V NMR spectra of sample S0 obtained at $T=380$~mK
for different applied magnetic fields in the range $40.45 < \mu_0 H < 45$~T.
Inset: Data for sample S2 are plotted additionally (solid line).
The base lines of the different spectra in the inset are shifted
along the $y$--axis for better visibility.}
 \label{Figure-8}
\end{figure}

\begin{figure}
\includegraphics[width=80 mm]{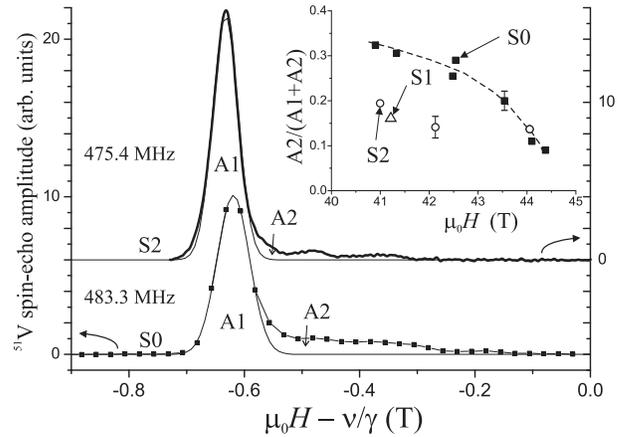}
\caption{$^{51}$V NMR spectra at $T=380$~mK for samples S0 and S2
measured at 483.3 and 475.4 MHz, respectively. Solid lines show
the fitting results of the main peak of the spectral pattern with a
Lorentzian line (A1). Inset: Relative integral intensity of the
residual contribution A2 of the NMR spectra pattern which does not
belong to the main line A1. Data from different samples S0 (solid
squares), S1 (open triangle), and S2 (open circles) are shown.} \label{Figure-9}
\end{figure}

We return to a closer inspection of the discrepancy between
the internal field at the majority of vanadium nuclei obtained from
NMR experiments and the bulk magnetization $M(H)$ observed within the
range of applied magnetic fields $H_{\rm c3} < H < H_{\rm sat}$.
The natural explanation for this phenomenon is the non--uniform
magnetization of the sample, i.e., the magnetization of the majority
part of the sample is saturated already at $\mu_0 H = 41.4$~T,
whereas there are local regions within the sample whose
magnetization is saturated only at $H_{\rm sat}$. In such a case,
the intense line of the NMR spectra in the high--field range is
ascribed to the nuclei surrounded by magnetic moments with the
saturated magnetization, whereas the experimentally observed
growth of the bulk magnetization $M(H)$ of all the samples can be
attributed to the non--saturated region near defects.

The expected magnetic structures for the field ranges $H_{\rm c2} <
H < H_{\rm c3}$, $H_{\rm c3} < H < H_{\rm sat}$, and $H>H_{\rm sat}$
are illustrated in Figs.~\ref{Figure-7}~A, C, and D, respectively.
The structures in panel A (below $H_{\rm c3}$) and C (above $H_{\rm
c3}$) both exhibit longitudinal modulation of the magnetization. In
case of the spin--modulated phase (A) this modulation has a regular
sinusoidal shape, whereas for the high--field phase (C) the
modulation is not periodic but accidental. We expect that the latter
structure is associated with crystallographic defects. Although the
nature of the defects is not precisely known, a
composition study of LiCuVO$_4$ indicates the presence of Li defects of
a few percents even in carefully prepared
crystals.\cite{Prokofiev_04} It was argued that a Li deficiency
produces a hole doped into the oxygen sites, which in turn will form
a Zhang--Rice singlet with a Cu spin.\cite{Prokofiev_04,Zhang_88} Such
a singlet should be equivalent to a non--magnetic defect replacing
a Cu spin and provokes an unusual magnetic state in its vicinity
with deleted two nearest ferromagnetic exchange bonds, but with a
conserved antiferromagnetic bond between the two parts of the interrupted chain
(see panel B of Fig. \ref{Figure-7}). A theoretical study shows that
the saturation field of the magnetic neighborhood of such
crystallographic defects is expected at distinctly higher fields
compared to the saturation field of undisturbed magnetic
chains.\cite{Zhitomirsky_13}

In the following we describe the fingerprints of such defect structure
found in our experiments and evaluate the concentration of those defects.
Figure~\ref{Figure-8} is a replot of the field swept $^{51}$V NMR
spectra for applied magnetic fields $H$ between $40.45 < \mu_0 H <
45$~T at 380~mK. Here, the ordinate axis is adjusted for each
spectrum in such a way that the integrated intensity of the entire
line is the same for all the spectra. This representation allows for
an estimation of the defect concentration. The NMR spectra in this
field range consist of an intense symmetric line and a broad
plateau--like part. The latter component is closer to the
diamagnetic reference field at $\mu_0 H-\nu/\gamma = 0$, and hence has
smaller internal fields. The intense symmetric line can be fit by a
Lorentzian denoted A1 in Fig.~\ref{Figure-9}, allowing us to
separate the plateau--like part marked as A2. The inset of Fig.~\ref{Figure-9}
shows the field dependence of the relative weight of
the plateau--like part A2/(A1+A2). For sample S0 this relative
intensity of the plateau--like part decreases monotonously within
the entire field range $H_{\rm c3}< H < H_{\rm sat}$, whereas for
the samples S1 and S2 it was found to be smaller and to exhibit a
weaker decrease with increasing field $H$.

The relative intensity A2/(A1+A2) at fields in the vicinity of $H_{\rm sat}$ is
approximately 0.1. From this value we can
evaluate the defect concentration $x$: the probability $p$ that all four
magnetic sites surrounding the $^{51}$V nucleus in its nearest proximity are
occupied by Cu$^{2+}$ ions is equal to $p = (1-x)^4$. Here we suppose that
defects are distributed randomly. On the other hand, this value can be
evaluated as $A1/(A1+A2)\approx 0.9$. From this equation we obtain the
concentration $x = (2.5\pm 0.6) \%$ of non--magnetic defects in reasonable
agreement with the detailed investigation of reference~\onlinecite{Prokofiev_04}.

\section{Conclusion}
In conclusion, we investigated the magnetic structure of the
Cu$^{2+}$ moments in the quantum--spin chains of LiCuVO$_4$ in
applied magnetic fields $H$ up to the saturation field $\mu_0H_{\rm
sat} = 45$~T. The double--horn shape of the $^{51}$V NMR spectra in
the incommensurate SDW phase changes to a single Lorentzian line
around $\mu_0H_{\rm c3}=(40.5 \pm 0.2)$~T. Although the magnetization
curve $M(H)$ shows a linear increase of magnetization in the field range
$H_{\rm c3} < H < H_{\rm sat}$,\cite{Svistov_11} the
internal field corresponding to the peak of the NMR spectra stays
constant for $\mu_0 H > 41.4$~T, indicating that the moments
surrounding the majority of vanadium nuclei are saturated in this field
range. The results of the nuclear spin--lattice relaxation rate of $^{7}$Li nuclei
also show an energy gap expected for bound magnon pairs above the
saturation at $\mu_0 H \approx 41.4$~T. From these results, we
conclude that the theoretically predicted nematic ordered phase can
be realized only in the narrow field range  $\mu_0 H_{\rm c3} <
\mu_0 H < 41.4$~T if it exists in LiCuVO$_4$.

We have attributed the discrepancy between our NMR and magnetization
data to effects caused by defects. From the careful investigation of
a small pedestal--like contribution in the $^{51}$V NMR spectra,
we were able to quantify the number of magnetic defects.
As these defects locally yield much higher saturation
fields $H_{\rm sat}$ we reconciled our NMR results with recent
magnetization $M(H)$ experiments.\cite{Svistov_11}
Whether or not elimination of the defects stabilizes the spin--nematic phase
is an interesting future issue.

\begin{acknowledgments}

This work is supported by the German Research Society
(DFG) within the Transregional Collaborative Research Center TRR 80 (Augsburg, Munich),
by the Grants 13-02-00637 of the Russian Foundation for Basic Research, Program
of Russian Scientific Schools, by a Grant-in-Aid for Scientific Research 25287083 from the Japan Society of
Promotion of Science, and by the Global CEO Program (Core
Research and Engineering of Advanced Materials-Interdisciplinary
Education Center for Materials Science)(No.~G10) from MEXT, Japan.
Work at the National High Magnetic Field Laboratory is supported by
the NSF Cooperative Agreement No. DMR--0654118, and by the State of Florida.
We thank H. Tsunetsugu, N. Shannon, A. Smerald, M. Sato, T. Momoi,
T. Hikihara, H. T. Ueda, A. Matsuo, A. Smirnov, S. Sosin, O. Starykh, and M. Zhitomirsky
for stimulating discussions.

\end{acknowledgments}

\end{document}